\documentstyle[12pt]{l-aa}

\begin{document}

   \thesaurus{
              (03.20.4;  % Techniques: photometric
               08.08.1;  % Stars: HR-Diagrams
               08.12.3;  % Stars: Luminosity Function
               11.09.5;  % Galaxies: Irregular
               11.19.5)} % Galaxies: Stellar Content
   \title{The stellar content of the Sagittarius Dwarf Galaxy
\thanks{Based on observations collected at the
    European Southern Observatory, La Silla, Chile.}}

   \author{G. Marconi$^1$, R. Buonanno$^1$, M. Castellani$^1,^2$,
G. Iannicola$^1$, P. Molaro$^3$, L. Pasquini$^4$, L. Pulone$^1$
          }

   \offprints{G. Marconi}
   \institute{$^1$ Osservatorio Astronomico di Roma \\ 
Via dell'Osservatorio 2, Monte Porzio Catone, 00040 (Roma). Italy \\
{\it [marconi,buonanno,mkast,iannicola,pulone]@coma.mporzio.astro.it} \\
$^2$ Universita' "La Sapienza", Istituto Astronomico \\ 
Via Lancisi 29, 00161 Roma \\
$^3$ Osservatorio Astronomico di Trieste,
Via G.B. Tiepolo 11, 34131 Trieste  \\
$^4$ ESO, La Silla}

   \date{Received September; accepted}

   \maketitle

\begin{abstract}
We present V,I deep
CCD photometry for three fields of the 
dwarf galaxy in Sagittarius (Sgr), located at l=5.6$^o$, b=-14.1$^o$.
One of the fields is 
centered on the globular cluster NGC 6715 (M54), 
which lies in one of the dense clumps of the Sgr  galaxy.
Comparing the CMD of Sgr with those of 
globular clusters which are believed to be kinematically 
associated with the dwarf galaxy (Da Costa \&
Armandroff, 1995), 
we conclude that the stellar population of
Sgr presents a spread in metallicity of   
-0.71$\leq$[Fe/H]$\leq$-1.58, and that the dominant population 
($\simeq$10 Gyr old) is extremely similar to the star content of 
the associated globular cluster Terzan 7. The estimated distance to Sgr is
d$\simeq$24.55 Kpc.

\keywords{techniques: photometry - stars: HR-diagrams - stars: luminosity 
function - galaxies: irregular - galaxies: stellar content} 

\end{abstract}

%%%%%%%%%%%%%%%%%%%%%%%%%%%%%%%%%%%%%%%%%%%%%%%%%%%%%%%

\section{Introduction}
\par
Ibata, Gilmore and Irwin (1994, 1995; hereafter both indicated with IGI) 
identified a new dwarf galaxy 
of the Local Group, 
located in the 
constellation of Sagittarius. The dwarf, detected as a 
moving group of stars with mean
radial velocity significatively different from that of the stars in the bulge,
is located 
at a distance of about 25 Kpc from the Sun, and
subtends an angle of $\simeq$10 degrees on the sky. The new galaxy is
comparable in size and luminosity to the dwarf
spheroidal Fornax, which is the largest galaxy
known in the Local Group. The recent discovery of two planetary nebulae
(Zijlstra \& Walsh, 1996) confirms that Sgr is similar in mass to the Fornax
dwarf, which contains a planetary nebula as well.
The other, less
massive, dwarfs of the Local Group 
seem not to 
contain any star in such a
short evolutionary phase.

Although the 
CM diagram of IGI is heavily contaminated by the 
rich population of foreground
disk and bulge stars, nonetheless it clearly shows the presence of an 
intermediate-old dominant population. 

Mateo et al. 1995a (hereafter MUSKKK), presented
the CCD photometry of a field
in Sagittarius, which reaches 
I$\simeq$22.3. After having statistically removed the field
stars from the CM diagram, they concluded that Sgr is dominated by a
intermediate-old population which is younger than that of the bulk 
of galactic  globular clusters. They also found evidence of
a feature in the CMD which could be interpreted as
a weak component of intermediate age or, alternatively, as a population of
blue stragglers. Finally, from the color of the
red giant branch they estimated that the 
metallicity of the dominant
population is about [Fe/H]$\simeq$-1.1. 
Mateo et al. (1995b) extensively discussed the properties of a handful of 
RR Lyrae variables in the field of Sgr.

Sarajedini \& Layden (1995, hereafter SL) presented the V,I CCD color
magnitude diagram for two fields. The first field is 
centered near the globular cluster M54
and covers both the cluster 
and the central region of Sgr, while the second field is  
centered 12 arcmin north of the cluster.
From the analysis of the CMD 
they found that 
the bulk of Sgr is formed by a relatively metal-rich population of
[Fe/H]$\simeq$-0.52, and that  
a component of lower metallicity ([Fe/H]$\simeq$-1.3) could also
be present.

Sgr is the nearest galaxy to the
Milky Way
which has been discovered until now, and
numerical simulations confirm that Sgr is presently in
phase of being disrupted by the Galaxy 
(Kathryn, Spergel \& Hernquist 1995).

This idea is supported by Mateo et al. (1996) 
and by Alcock et al. (1997), who found 
that Sgr is considerably more
extended than previously believed, 
suggesting 
that this 
dwarf has
probably suffered at least one - and perhaps many - strong tidal
encounters with the Milky Way during past perigalacticon passages.

Another piece of evidence of the strong interaction between Sgr and the 
Milky Way is that two of the clusters (Ter 7 and Arp2) 
which have radial velocities similar to that
of Sgr, 
and then suspected to be 
cinematically associated with Sgr (Da Costa and Armandroff, 1995),
are anomalously young compared to the bulk of galactic globulars
(Buonanno et al. 1994, 1995a,b).

Recently
Fahlman et al. (1996; hereinafter FMRTS) 
detected the upper end of the main sequence
in the background field around the globular cluster M55, which is 
projected in the sky about 40 arcmin east of Sgr.
Having interpreted this 
feature as 
the turn-off of Sgr, they used the Vandenberg \& Bell 
(1985) isochrones to give
an estimate of the age of the galaxy,
and concluded that 
Sgr is between 10 and 13 Gyr old, although this conclusion is
admittedly based on an handful of stars.

Finally, the
complex history of Sgr has been confirmed by
Ng \& Schulthesis (1996) who detected a carbon-star which,
if its membership to Sgr will be confirmed,
hints at the existence of a stellar population about 4 Gyr younger
than the dominant population.

In this paper we present deep V,I CCD 
photometry of Sgr, in two separate fields, far from those 
of SL and MUSKKK. Our aim is
twofold: first, to explore whether the results above summarized 
depend on the specific region of Sgr examined (considering the 
extension of Sgr this aspect is 
particularly important) and, second, to 
investigate the nature of the Sgr population. 

Deep V,I photometry of the field near M54, already studied by SL, is also presented.

In section 2, we describe the observations 
and the reduction procedures, 
while the resulting C-M diagrams are presented and discussed in section 3. The 
stellar content of Sgr is studied and compared to that 
of associated globular clusters in 
section 4. 
The conclusions are drawn in section 5.

%%%%%%%%%%%%%%%%%%%%%%%%%%%% OSSERVAZIONI %%%%%%%%%%%%%%%%%%%%%%%%%%%%%%%%%%

\section{Observations and reductions} 
\par 
\subsection{Observations} 
\par
Observations of a field which includes a portion of Sgr and of M54
($\alpha$$_{2000}$=18$^h$55'03'', $\delta$$_{2000}$=-30$^o$28'42'')
were first obtained using EFOSC2 at the ESO/MPI 2.2 m at La Silla
during September 1994. 

In August 1995 additional
observations of Sgr were obtained with 
the 3.5 m ESO-NTT. The telescope was equipped with EMMI
which mounted 
the coated Loral CCD 2048 X 2048.  
%at the Nasmyth focus A, with a scale of 0.12" per pixel.  
A series of
frames was taken
in two different field of the galaxy,  
one centered at $\alpha$$_{2000}$=18$^h$53'44'', $\delta$$_{2000}$=-30$^o$28'23''
% "north" of M54   
({\it field 1}) and the other centered at  
$\alpha$$_{2000}$=18$^h$59'44'', $\delta$=-30$^o$59'45''
({\it field 2}). 

The three observed fields are 
sketched in figure 1 (a,b,c).  
Several  Landolt  (1992) stars (from 7 to 15), 
were observed for calibration throughout each night.

The details of the observations, with 
the exposure time 
and seeing conditions, are listed in Table 1. 

\begin {table}
\caption[]{Journal of observations}
\begin{flushleft}
\begin{tabular}{rrrrrrl}
\ \\
\hline
\hline
\ \\
$Date $&$Filter $&$Exp. time$&$Seeing$&$Field$&$Tel.$\\
\hline
\ \\
% 6 aug. 95 & B & 480 s & $0.77^{\prime\prime}$ & Sgr 1 & NTT \\
6 aug. 95 & V & 240 s & $0.70^{\prime\prime}$ & Sgr 1 & NTT \\
6 aug. 95 & I & 300 s & $0.74^{\prime\prime}$ & Sgr 1 & NTT \\
% 7 aug. 95 & B & 480 s & $0.72^{\prime\prime}$ & Sgr 1 & NTT \\
% 7 aug. 95 & B &  40 s & $1.00^{\prime\prime}$ & Sgr 1 & NTT \\
7 aug. 95 & V &  20 s & $1.12^{\prime\prime}$ & Sgr 1 & NTT \\
7 aug. 95 & I &  30 s & $0.95^{\prime\prime}$ & Sgr 1 & NTT \\
6 aug. 95 & V & 240 s & $0.89^{\prime\prime}$ & Sgr 2 & NTT \\
6 aug. 95 & I & 300 s & $0.67^{\prime\prime}$ & Sgr 2 & NTT \\
7 aug. 95 & I & 300 s & $0.65^{\prime\prime}$ & Sgr 2 & NTT \\
% 7 aug. 95 & B & 480 s & $0.62^{\prime\prime}$ & Sgr 2 & NTT \\
% 7 aug. 95 & B &  40 s & $0.96^{\prime\prime}$ & Sgr 2 & NTT \\
7 aug. 95 & V &  20 s & $0.80^{\prime\prime}$ & Sgr 2 & NTT \\
7 aug. 95 & I &  30 s & $0.86^{\prime\prime}$ & Sgr 2 & NTT \\
7 sep. 94 & B &  60 s & $0.85^{\prime\prime}$ & M54 & 2.2 \\
7 sep. 94 & B &  1200 s & $0.93^{\prime\prime}$ & M54 & 2.2 \\
7 sep. 94 & B &  600 s & $0.98^{\prime\prime}$ & M54 & 2.2 \\
7 sep. 94 & B &  300 s & $0.95^{\prime\prime}$ & M54 & 2.2 \\
7 sep. 94 & V &  300 s & $0.89^{\prime\prime}$ & M54 & 2.2 \\
7 sep. 94 & V &  300 s & $0.92^{\prime\prime}$ & M54 & 2.2 \\
7 sep. 94 & V &  240 s & $0.95^{\prime\prime}$ & M54 & 2.2 \\
7 sep. 94 & V &  240 s & $0.95^{\prime\prime}$ & M54 & 2.2 \\
7 sep. 94 & V &  120 s & $0.92^{\prime\prime}$ & M54 & 2.2 \\
7 sep. 94 & I &  300 s & $0.88^{\prime\prime}$ & M54 & 2.2 \\
8 sep. 94 & I &  180 s & $0.82^{\prime\prime}$ & M54 & 2.2 \\
8 sep. 94 & B &  300 s & $0.90^{\prime\prime}$ & M54 & 2.2 \\
8 sep. 94 & B &  300 s & $0.99^{\prime\prime}$ & M54 & 2.2 \\
8 sep. 94 & B &  480 s & $0.95^{\prime\prime}$ & M54 & 2.2 \\
8 sep. 94 & I &  300 s & $0.89^{\prime\prime}$ & M54 & 2.2 \\
8 sep. 94 & I &  180 s & $0.91^{\prime\prime}$ & M54 & 2.2 \\
8 sep. 94 & I &  180 s & $0.93^{\prime\prime}$ & M54 & 2.2 \\
8 sep. 94 & I &  180 s & $0.82^{\prime\prime}$ & M54 & 2.2 \\
8 sep. 94 & I &  180 s & $0.85^{\prime\prime}$ & M54 & 2.2 \\
8 sep. 94 & I &  180 s & $0.85^{\prime\prime}$ & M54 & 2.2 \\
8 sep. 94 & I &   60 s & $0.86^{\prime\prime}$ & M54 & 2.2 \\
8 sep. 94 & I &   60 s & $0.88^{\prime\prime}$ & M54 & 2.2 \\
8 sep. 94 & I &   60 s & $0.93^{\prime\prime}$ & M54 & 2.2 \\
\ \\
\hline
\end{tabular}
\end{flushleft}
\end{table}

%%%%%%%%%%%%%%%%%%%%%%% FIG 1a-1b-1c %%%%%%%%%%%%%%%%%%%%%%%%%%%%%%%%%%

\begin{figure}[ht]
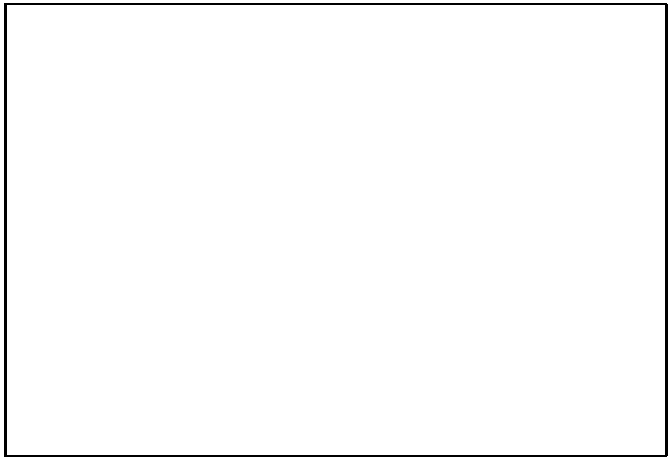

\picplace{6cm}
\caption[]{The Sagittarius Dwarf (a=field 1, b=field2, c=field M54)}
\end{figure}

%%%%%%%%%%%%%%%%%%%%%%%%%%%%%%%%%%%%%%%%%%%%%%%%%%%%%%%%%

\subsection{Reduction Procedures}
\par
After having applied the basic procedures of pre-reduction 
(bias and dark current subtraction, trimming, flatfield
correction, etc.) we  
reduced the NTT frames using DAOPHOT II
(Stetson 1991), while the frames taken at the 2.2 m, 
which appeared extremely crowded, 
were reduced 
using ROMAFOT (Buonanno et al., 1983, Buonanno \& Iannicola, 1989),
because this package allows an easy and continuos interaction of
the operator during the reductions.  
 
We detected 9154 stars in {\it field 1} and 6975 stars in {\it field 2} 
down to V$\simeq$23 
for the NTT observations and 
7928 stars in the field around M54,
observed with the 2.2 m.

The tables giving the x,y coordinates, relative to an arbitrary origin, 
V magnitudes and V-I colors, can be obtained via {\it Internet} from host
{\it "coma.mporzio.astro.it"} using the "anonymous ftp" service (directory
{\it /pub/SAGIT}; files {\it sagit1.cal}, {\it sagit2.cal}, {\it M54.cal} and
{\it README}.
The conversion from  instrumental magnitudes into the Johnson standard system 
was obtained using a set of primary calibrators (Landolt, 1992)
which spanned a wide range in color
($-0.360\leq$V-I$\leq2.030$), and 
has been computed separately for each 
telescope and for each photometric night. 

For NTT we obtained  color equations in the form: 

\smallskip\noindent
V = v - 0.001($\pm$ 0.020)$\cdot$ 
(v-i) + {\it const.}

\smallskip\noindent
I = i - 0.076($\pm$ 0.020)$\cdot$ 
(v-i) + {\it const.}

\smallskip\noindent
While for the 2.2m we obtained: 

\smallskip\noindent
B = b + 0.348($\pm$ 0.010)$\cdot$ 
(b-v) + {\it const.}

\smallskip\noindent
V = v + 0.096($\pm$ 0.005)$\cdot$ 
(v-i) + {\it const.} \smallskip\noindent

\smallskip\noindent
I = i + {\it const.} \noindent

where B, V and I are the magnitudes in 
the standard system, and b, v and i are 
the instrumental magnitudes. The formal errors in the zero 
point of the calibration were: 0.04 mag in V and 0.03 mag in I
for the NTT data;  
0.05 mag in B, 0.03 mag in V and 0.02 mag  in I for the 2.2m data. 
The r.m.s. of the
residuals for the standards, respectively 
for the NTT and the 2.2 m, resulted
$\sigma_{V_{NTT}}$=0.018, $\sigma_{I_{NTT}}$=0.012, 
$\sigma_{B_{2.2}}$=0.022,
$\sigma_{V_{2.2}}$=0.013, $\sigma_{I_{2.2}}$=0.009.

%%%%%%%%%%%%%% DIAGRAMMA CM %%%%%%%%%%%%%%%%%%%%%%%%%%%%%%%%%%%%%%%%%

\section{Color magnitude diagrams for Sgr and M54}
\par

Figures 2 and 3 show the V,  V-I diagrams for the stars of Sgr, 
respectively in {\it field 1} and {\it field 2}, while the CMD of the field 
centered on M54 observed at the 2.2m is presented in figure 4a for
the whole frame excluding the most central region, (r$\leq$60 arcsec), 
and in figure 4b for the annulus with 134$\leq$r$\leq$157 arcsec.
The latter region, 
selected as it presents both a  
manageable degree of crowding and a high probability of
cluster membership, allows a reliable estimate of the turn-off of M54.

%%%%%%%%%%%%%%%%% FIG 2 %%%%%%%%%%%%%%%%%%%%%%%%%%%%%%%%%%%%%%%%%%%%%%%%%

\begin{figure}[ht]
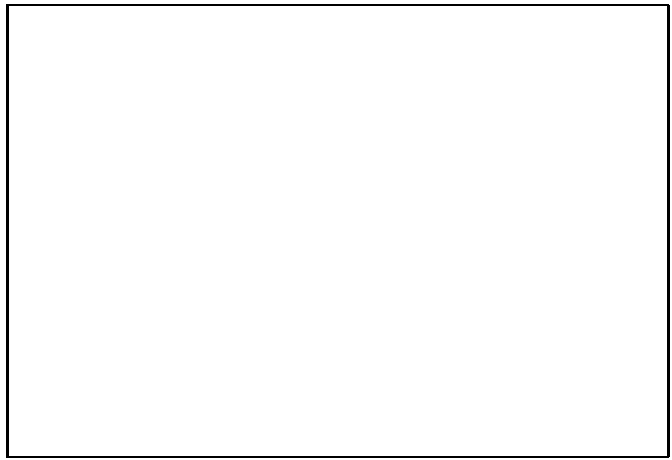

\picplace{6cm}
\caption[]{CMD of stars in Sagittarius field 1. The error bars for bins
of magnitude and colour are shown on the left of the diagram}
\end{figure}

%%%%%%%%%%%% FIG 3 %%%%%%%%%%%%%%%%%%%%%%%%%%%%%%%%%%%%%%%%%%%%%%%%%%%%%%%%%

\begin{figure}[ht]
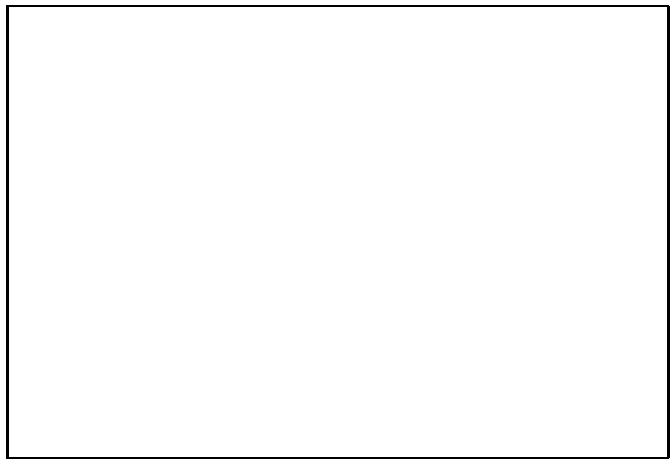

\picplace{6cm}
\caption[]{CMD of stars in Sagittarius field 2.}

\end{figure}

%%%%%%%%%%%%%%%%%%%%%%%%%%%%%%%%%%%%%%%%%%%%%%%%%%%%%%%%%%%%%%%%%%%%%%

Two major features to note in 
figure 4a are, first, a well populated, steep RGB with V$_{tip}\simeq$15.2
and $(V-I)_{tip}\simeq$1.75
and, second, a well-defined Blue Horizontal Branch (BHB), located at  
V$\simeq$18.2 and extending towards (V-I)$\simeq$0.
Since both these features are clearly 
absent in figures 2 and 3, we confirm  
the suggestion of 
SL that they must belong to M54.

Figure 4a also shows another fairly-populated  sequence
which  emerges 
near V$\simeq$19.5 and, extending towards the bright red side
of the CMD, reaches V$\simeq$16.2 and  (V-I)$\simeq$1.9.
This sequence, easily detectable also in figg. 2 and 3, 
has been identified by SL as the RGB of the dominant 
population of Sgr.

%%%%%%%%%%%%%%%% FIG 4 %%%%%%%%%%%%%%%%%%%

\begin{figure}[ht]
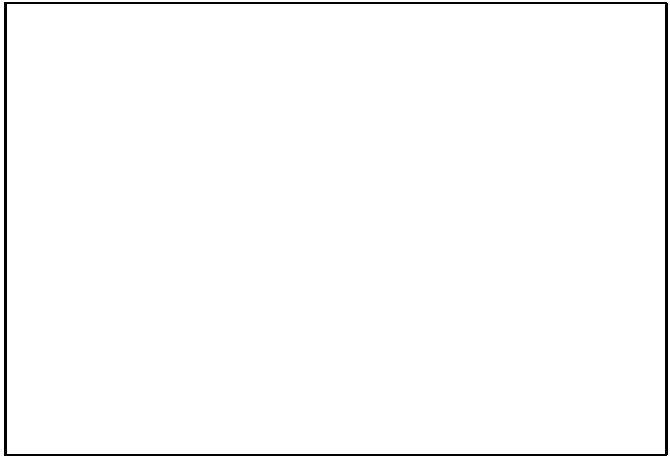

\picplace{6cm}
\caption[]{(a) CMD  of the  field 
centered on M54, (a) for the region r $\geq$ 60'' 
(a) and,  for the annulus with  
134''$\leq$ r $\leq$ 157'' (b)}

\end{figure}

%%%%%%%%%%%%%%%%%%%%%%%%%%%%%%%%%%%%%%%%%

Passing to the CMDs of {\it field 1} and {\it field 2} we note, first of all,
that they are extremely similar, witnesses of a homogeneous evolution
of this dwarf galaxy.

The most prominent features detectable in figures 2 and 3 are:

a) a clump of stars (V $\simeq$18.1 and 
V-I $\simeq$1.1) located
at the blu side of the red giant branch, already identified 
by MUSKKK and SL as the HB locus of the dominant population of Sgr
(such feature is much more evident in figure 2).
The same clump 
is also visible in figure 4a (V $\simeq$18.25),  
where it appears embedded in the rich RGB of M54, 
but slightly shifted to the red. 

b) a fairly steep RGB which reaches V$\simeq$15.5 and 
V-I$\simeq$1.9 whose observed dispersion, much larger than  the
photometric errors, hints at a dispersion in metallicity 
within Sgr.
Such RGB is the most prominent feature of Sgr,
and is well visible also in the field centered on 
M54 in fig. 4a.
A closer inspection of figure 4a, however, reveals that the metal-rich 
RGB in the field of M54, measured at its base, 
is redder by {$\Delta$ (V-I) = 0.12$\pm$0.06 
than the RGB in {\it fields 1 and field 2}. 

There are three possibilities to explain this observational result:
first, an
inconsistency of  the two (independent) calibrations; second, a differential 
reddening within Sgr;  
third, a metallicity gradient within Sgr. 
To disentangle the problem, we first compared the RGB loci of the CMD in
{\it field 1} and {\it field 2} with the data of MUSKKK (their fig. 1).
This comparison, shown  in figure 5, discloses that
a good agreement exists between the data 
of MUSKKK and those presented in this paper, supporting the 
reliability of  our
 calibration of {\it field 1} and {\it 2}. 

Then we compared the data of figure 4 with those of SL 
for the field centered on M54.
Figures 6a and 6b show the differences 
in color and magnitude for the 1295 stars in common.
Both the diagrams of fig. 6 show a fairly simmetrical distributions  
around the zero, with a slight residual color equation, of the
order of 0.04 mag, for (V-I)$\simeq$1.5. Since this color equation 
goes in the sense that the RGB of SL is 
 even redder than
that obtained in the present paper for M54,
we conclude that the difference detected in the RGB colors 
 cannot be attributed to uncertainties in the calibration of the three 
fields of Sgr.
We are inclined, therefore, to infer that 
the stars of Sgr present an 
intrinsic color gradient within the main body of the dwarf.
  
We pass now to examine whether 
the galactic reddening is responsible of the observed color shift 
of $\Delta$(V-I) = 0.12$\pm$0.06. If this would be the case, 
 we should expect to observe that the red HB clump,
would result in the  {\it field M54}
fainter   
than in the {\it fields 1} and {\it 2} by $\Delta$V = 0.30$\pm$0.15. 
 Actually we observed that the 
red HB clump in the {\it field M54} is fainter 
than the clumps in {\it fields 1} and  {\it 2}  
by only $\Delta$V = 0.15$\pm$ 0.03 magnitudes, which is marginally 
consistent with the value of $\Delta$V=0.30$\pm$0.15 expected under the 
hypotesis that a absorption gradient exists whithin Sgr.  

We therefore remain   
with the last possibility, i.e. that the observed  color-shift is due 
to a spread in metallicity. To check the viability of such hypothesis,
we estimated the possible spread in metallicity using 
the empirical relation of  Castellani et al. (1996)
$\Delta$[Fe/H] = $\Delta$(V-I) / 0.24. From  
$\Delta$(V-I) = 0.12$\pm$ 0.06 
 one obtains   $\Delta$[Fe/H] =0.50$\pm$0.25. 
To see how such difference in metallicity would reflect in the luminosity of 
the HB clump, we used the relation of Lee et al. (1990) 
$M_V$(HB) = 0.15$\cdot$[Fe/H]+0.82, and obtained $\Delta$V = 0.09$\pm$0.04, 
which is in good agreement with the obseved value $\Delta$V = 0.15$\pm$ 0.03.
 It seems therefore, that a difference in metallicity 
of $\Delta$[Fe/H]$\simeq$0.50 accounts for all the photometric observables of 
{\it field 1} {\it field 2} and {\it field M54} of Sgr.  

There is, actually,  an  alternative explanation based on the ambiguity 
in the back-to-front ratio in Sgr. Interpreting the 
 location of the RGBs in  different  
 fields of Sgr as a difference in magnitude
 instead of a difference in color, one can slide the CMD of {\it field M54} by 
about 0.15 mag brighter and superimpose it to the CMD of 
{\it fields 1} and {\it 2}.
 In other words the observed features are compatible with a bizarre shape 
of Sgr which would present the two edges ({\it fields 1} and {\it 2}) 2.3 
Kpc nearer to the 
Sun than the central region ({\it field M54}).

c) A  sequence which runs almost vertically, starting at 
V-I$\simeq$0.8 and extending up to V$\simeq$14.0.
It tends to become redder for fainter 
magnitudes, showing the typical shape of a sequence of bulge stars.
Although the faint extension of the vertical sequence  
makes the Sgr turn-off
region somewhat confused, the main structures are 
nevertheless clearly visible, allowing a rough estimate of the location
of the turn-off point.
To this purpose we first splitted the cumulative CMDs of 
{\it fields 1} and {\it 2}
in intervals of 0.2 magnitudes, then for each bin 
we computed the histograms in 
color of the stars fainter than V = 20.5. Finally, in order to 
determine the mode of the color histogram, we 
performed a gaussian fit to each distribution 
whose modal values are reported in table 2. Inspection of table 2 
reveals that the 
bluest color of the ridge line is V-I $\simeq$ 0.70, corresponding to the 
magnitude interval centered at V = 21.2. Therefore, we will assume, as 
a first guess, $V_{TO}$ = 21.2 $\pm$ 0.1 for Sgr.

d) A nearly horizontal sequence,
which starts
from the position of the Sgr HB clump, 
and reaches V-I$\simeq$0.7, crossing the vertical sequence of 
field stars.

\begin {table}
\caption[]{The ridge line of Sgr for the turn-off region}
\begin{flushleft}
\begin{tabular}{ccc}
\ \\
\hline
\hline
\ \\
$V$&$(V-I)$$_{mode}$\\
\ \\
\hline
\hline
\ \\
20.6 & 0.81 & \\
20.8 & 0.77 & \\
21.0 & 0.73 & \\
21.2 & 0.70 & \\
21.4 & 0.71 & \\
21.6 & 0.72 & \\
21.8 & 0.75 & \\
22.0 & 0.76 & \\

\ \\
\hline
\end{tabular}
\end{flushleft}
\end{table}

%%%%%%%%% FIG 5 (NOI-MUSKKK) %%%%%%%

\begin{figure}[ht]
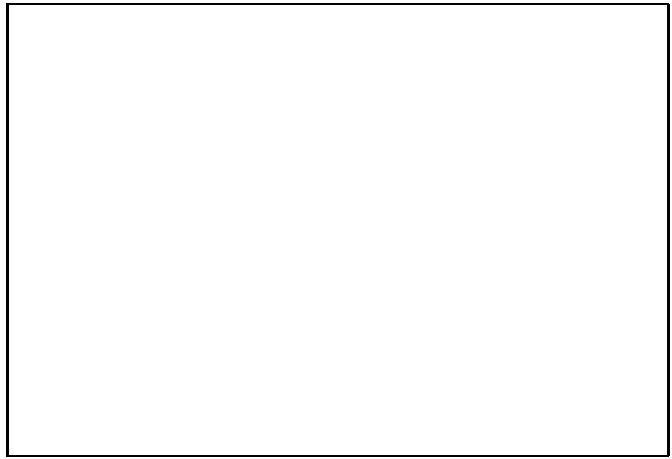

\picplace{6cm}

\caption[]{The RGB of Sgr. 
The data of figures 2 and 3 are indicated with full dots. The data of 
MUSKKK are reported as open dots. The good coincidence of the two set of data 
is clearly visible.}

\end{figure}

%%%%%%%%%%%%%%%%%%%%%%%%

%%%%%%%%%%% FIG 6 (CALIBR. SL vs. NOI) %%%%%%%%%%%%%

\begin{figure}[ht]
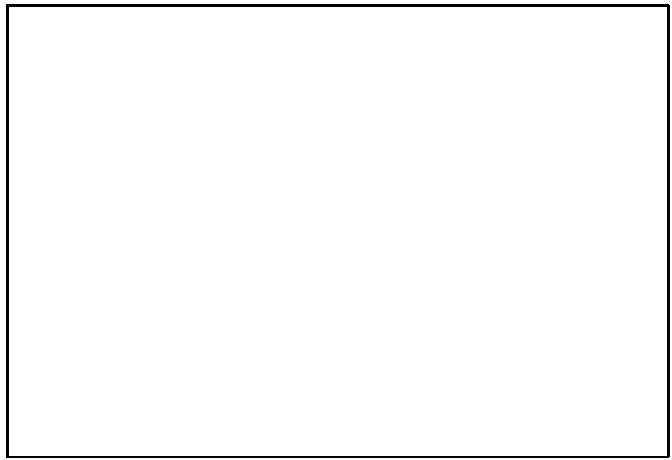

\picplace{6cm}
\caption[]{Star-by-star comparison of the present photometry 
({\it field M54}), with that of SL.}
\end{figure}

%%%%%%%%%%%%%%%%%%%%%%%%%%%%%%%%%%%%%%%%%%%%%%%%%%%

The magnitude level of this sequence, which appears ubiquitous  
in Sgr, can be derived from
the luminosity function 
of the stars detected in {\it field 1} in 
the interval 17.5$\leq$V$\leq$18.5 and V-I $\leq$ 1.0, 
here shown in figure 7.
Fitting a gaussian to the peack of the LF in the interval 
17.9$\leq$V$\leq$18.2, we derived $V_{peak}$ = 18.09 $\pm$ 0.08,
where the indetermination represents the halfwidth of the gaussian.

%%%%%%%%%%%%%%%%%%%%%

\begin{figure}
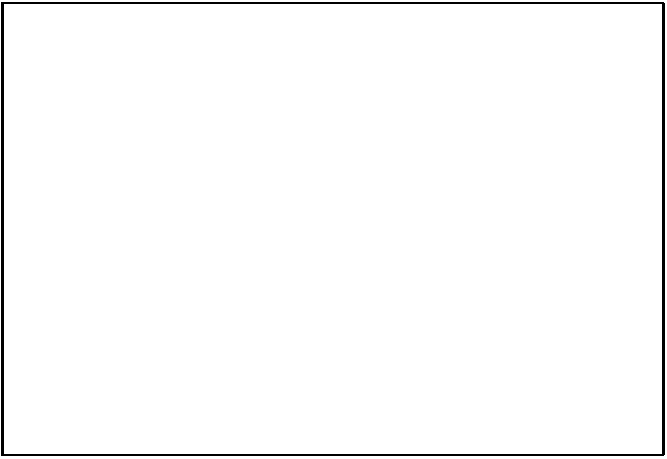


\picplace{6cm}

\caption[]{Histogram of the stars bluer than
V-I = 1.0; the location of the horizontal sequence it is easily seen at
$V\simeq 18.09$ (see text)}

\end{figure}

%%%%%%%%%%%%%%%%%%%%

In order to estimate the mean metallicity of the dominant population, 
we show in figure 8 the giant branch loci of the 
globular clusters M2 ([Fe/H]=-1.58) and 47 Tuc ([Fe/H]=-0.71) (data 
taken from Da Costa and Armandroff, 1990), overimposed to the 
CMD of Sgr {\it field 1} having adopted  $(m-M)_V$ = 17.50,
 $A_V$ = 0.55 and $E_{V-I}$ = 0.22 (see par. 4). Inspection of 
figure 8 reveals that the RGB of Sgr is delimited by the fiducial lines of 
M2 and  47 Tuc, indicating 
that the dominant population of Sgr falls in the range of 
metallicity -1.58$\leq$[Fe/H]$\leq$-0.71
The average value of metallicity, therefore turns out to be 
[Fe/H]=-1.1$\pm$0.3, in good agreement wiht the value found by MUSKKK
and at variance with that of SL.

%%%%%%%%%%%%%FIG 7 (TER7-SGR) %%%%%%%%%%%%%%%%%%%%%%%%%%%%%%%%%%%%%%%%%%%%%%%%%%%%%%%%%%%

\begin{figure}
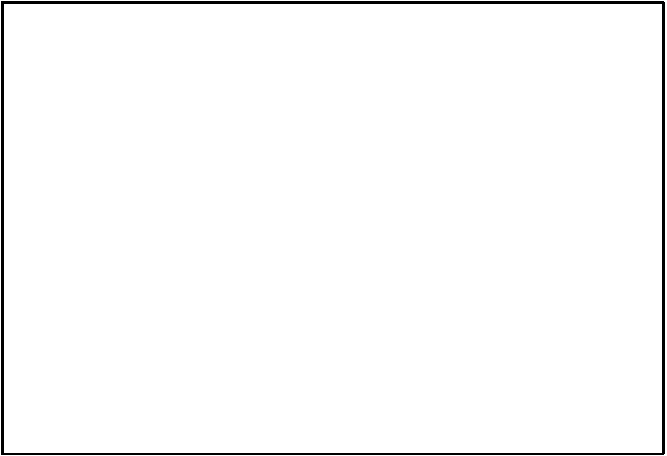


\picplace{6cm}

\caption[]{CMD (corrected for reddening
and extinction) of field 1 of Sgr with overimposed the  
 RGBs of the two galactic globular clusters 47 Tuc ([Fe/H]$\simeq$-0.71), 
and M2 ([Fe/H]$\simeq$-1.58)}

\end{figure} 

%%%%%%%%%%%%%%%%%%%%%%%%%%%%%%%%%%%%%%%%%%%%%%%%%%%%%%%%%%%%%%%%%%%%%%%%%

%%% da qui ricomicia la versione corretta .... %%%%

%%%%% SECT 4  
%%%%%%%%%%%%%%%%%%%%%%%%%%%%%%%%%%%%%%%%%%%%%%%%%%%%%%%%%%%%%%%%%%%%%%%%%%

\section{Stellar populations in the Sagittarius dwarf}

In order to understand the observed stellar content of a complex
system, the common practice is to select CMDs of the simplest
well-studied stellar groups, in general globular clusters, and then to
identify in the complex systems those features which 
characterize the template. This procedure is particularly 
strightforward for Sgr because Da Costa \& Armandroff (1995), studying
the relative distances and motions,
suggested that the globular clusters Terzan 7, Terzan 8, Arp 2 and M54 
could have been originated from the Sgr galaxy itself.
It is particularly significative in this context that two of these
clusters, Ter 7 and Arp 2, have been found peculiarly young and, therefore,
suspected to be captured by the Milky Way (Whitelock et al. 1996).

\subsection{Terzan 7}

The V, B-V CMD of Ter 7 was obtained by
Buonanno et al. (1995b).  
The main results of that study are
that Ter 7 is a metal-rich globular cluster (-1$\leq$[Fe/H]$\leq$-0.5,
depending on the adopted metallicity indicator), located at about 25 Kpc
from the galactic center whose age is about 4 Gyr lower than the bulk of
other globular clusters.

To perform the comparison with the present photometry of Sgr, 
we first derived an empirical 
relationship to transform the B-V to V-I colors. 
The relation we obtained from the photometry 
of several globular clusters,  is \par
\smallskip

(V-I) = -0.105$\cdot$(B-V)$^2$ + 1.221$\cdot$(B-V) - 0.078 \par\noindent

\smallskip

(valid in the colour range
0.0 $\leq$ B-V $\leq$ 1.5). 
\smallskip

This empirical relation turned out to be in 
excellent agreement with that of Smecker-Hane et 
al. (1995, hereafter SH), considered 
that the largest difference does not exceede
0.1 mag in color in the interval of validity (see table 3).

%%%%%%%% tebella 3 %%%%%%%%%%%%%%%%%%%%

\begin {table}
\caption[]{The empirical relationship between (B-V) and  (V-I) }

\begin{flushleft}
\begin{tabular}{cccc}
\ \\
\hline
\hline
\ \\
$(B-V)_{ours}$&$(V-I)_{SH}$&$(V-I)_{ours}$ \
\ \\
\hline
\ \\
0.0 & 0.00 & -0.08 & \\
0.5 & 0.54 & 0.51 & \\
1.0 & 1.08 & 1.04 & \\
1.5 & 1.62 & 1.52 & \\
\ \\
\hline
\end{tabular}
\end{flushleft}
\end{table}

%%%%%%% fine tab. 3 %%%%%%%%%%%%%%%%%%%

After having trasformed the V, B-V diagram of Ter 7 into the corresponding
V, V-I diagram, we adopted E$_{B-V}$=0.06 (Buonanno et al. 1995b),
and then, E$_{V-I}$=1.24$\cdot$E$_{B-V}$=0.07 (Cardelli et al., 1988),
 A$_V$=3.3$\cdot$E$_{B-V}$=0.198 for Ter 7, 
and E$_{V-I}$=0.22
and A$_V$=0.55 for Sgr (following MUSKKK),   

Figure 9a shows the superimposition of the CMDs of Ter 7 
and Sgr, having considered  the relative reddening and extinction.     

%%%%%%%%%%%%%%%%%%

\begin{figure}
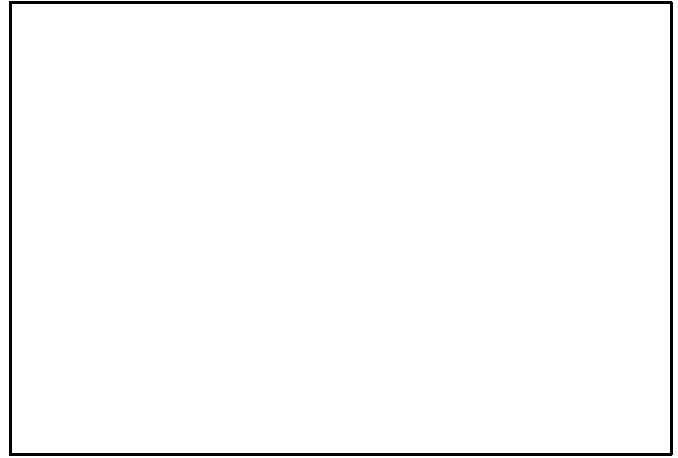


\picplace{6cm}

\caption[]{(a) superimposition of CMDs V vs V-I of Ter 7 
and Sgr (see text for details);
(b) CMD of Ter 7 with overimposed 
 the boxes used to count the stars along different evolutionary phases.}

\end{figure}

%%%%%%%%%%%%%%%%%%

It is of extreme interest to notice that the overall morphology of
Ter 7 reproduce remarkably 
well that
of Sgr in each evolutive phase. In particular 
the giant branches are exactly overlapped each other and the HB of Ter 7  lies 
in the same position of the HB clump of Sgr. 
In other terms, from a simple inspection of the relative location of the 
main branches, one concludes that Sgr lies at 
the same distance of 
Ter 7 from the sun, and that a population {\it Ter 7-like} is an
important contributor to Sgr.

The strong similarity of the two systems 
is reinforced noting 
that the TO of Ter 7 
is positioned exactly in the TO region of Sgr.
This conclusion can be put on more quantitative grounds:
having estimated, in fact,
V$_{TO}$$\simeq$21.2 (see  
paragraph 3),   
we obtained for Sagittarius  
V$_{TO_0}$=V$_{TO}$-0.55 = 20.65 $\pm$0.1, which is 
in excellent agreement with  V$_{TO_0}$ = V$_{TO}$-0.198=20.76$\pm$0.08 
obtained for Ter 7 (Buonanno et al. 1995b).   
All these evidences support the conclusion that Ter 7 has a common origin with 
Sgr, and that the age of the dominant population in Sgr is 
the same of that estimated by Buonanno et al. 
(1995b) for Ter 7, i.e. 4 Gyr lower than 47 Tuc. 

Given the importance of this issue, we ask now
whether the similarity between the dominant 
population of Sgr and that of Ter 7 
is confirmed by the lifetimes along the different 
evolutive phases. This can be accomplished by stellar counts along the 
different branches. Given  the crowding
along the branches of the CMDs, we first selected 
three boxes in which the counts 
  could be confidently performed. 
These boxes are sketched in figure 9b, where  
the separation between the different phases is well-defined.
The counts for the horizontal branch (labeled "HB"), 
the Red Giant Branch ("RG") and the 
Blue Stragglers region ("BS")
are reported in table 4, where columns 2 and 3 refer respectively to Sgr and
Ter 7, and column 4 shows the ratio of the counts for the two systems.
The errors 
are computed on the basis of the Poisson statistics. 

Inspection of table 4 
reveals that, altough one could expect to observe 
in Sgr a mixture of populations, the relative 
populations of Sgr and Ter 7 turn out to be extremely consistent,
leading to the conclusion that at least the 
portion of the galaxy sampled in this paper is largely constituted by the same 
stellar population of Ter 7. It is worth to notice that
the same conclusion holds for the population of blue 
stragglers, 
indicating that whatever are  
the parameters which favor the existence of BSs (primordial conditions, 
environment etc.), these parameters appear at work both in Ter 7 and in 
Sgr.

%%%%%%%%%%%%%%

\begin {table}
\caption[]{Stars in the various evolutive phases}
\begin{flushleft}
\begin{tabular}{ccccc}
\ \\
\hline
\hline
\ \\
$Phase$&$Sgr $&$Ter 7$&$Ratio$ \\
\hline
\ \\
HB & 97 & 24 & 4.04 $\pm$ 1.24 & \\
RG & 126 & 29 & 4.34 $\pm$ 1.19 & \\
BS & 61 & 14 & 4.36 $\pm$ 1.72 & \\
\ \\
\hline
\end{tabular}
\end{flushleft}
\end{table}

%%%%%%%%%%%%%%%

\subsection{Comparison with Arp 2}

The second comparison we performed is between Sgr and 
Arp 2 ([Fe/H]$\simeq$-1.8, Buonanno et 
al., 1995a). Figure 10 shows the CMD of Arp 2 superimposed to that of Sgr. 
For Arp 2 we adopted, following Buonanno et al. (1995b),
$E_{V-I}$=0.136 and $A_V$=0.34, while for Sgr we used the same color and
magnitude corrections of the 
previous paragraph. 
% To allow for the relative distance of Arp 2 and Sgr to the Sun,
An additional shift  
$\Delta$V=0.37 mag was finally applied, having adopted (m-M)$_0$=17.32 
and (m-M)$_0$=16.95 respectively for Arp 2 and Sgr.

%%%%%%%%%%%%%%

\begin{figure}
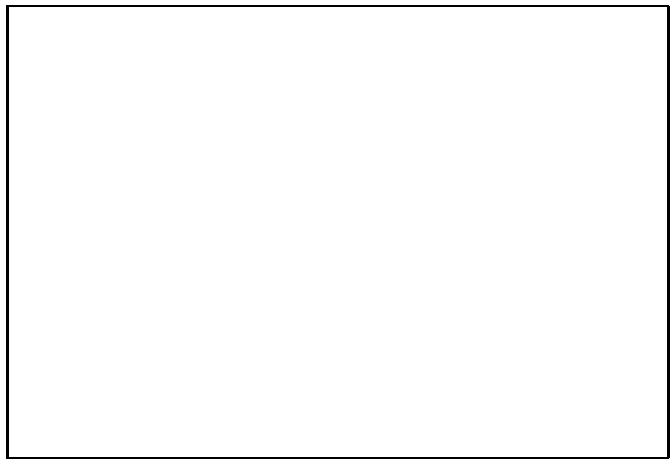


\picplace{6cm}

\caption[]{Superimposition of CMDs $V_0$ vs $(V-I)_0$ 
of Arp 2 and Sgr (see text for 
details)}

\end{figure}

%%%%%%%%%%%%%%

Inspection of figure 10 reveals that the giant branch of Arp 2 is located 
exactly in the region of the 
CMD of Sgr where several giants bluer than 
those of the dominant population of Sgr are observed. It would be 
therefore tempting to identify this 
region of the CMD as formed by a population 
{\it Arp 2-like}. There is however a significant 
obstacle to such identification, because Arp 2 presents 
a relatively rich population of 14 blu HB stars, 
which appears virtually absent in Sgr.

Considering that figure 10 suggests that the Sgr RGB we have temptatively
identified as {\it Arp 2-like} is grossly 
populated by the same number of stars of
the RGB of Arp 2, and considering that there are 14 HB stars in the CMD 
of Arp 2, one would expect to find 
of the order of 14 $\pm$ 4 HB stars
in Sgr.

A possibility, admittedly speculative, would be that the {\it Arp 2-like}
population of Sgr is slightly younger than Arp 2 itself by, say, about
2 Gyr.
This would shift the blue HB observed in Arp 2 to the red by 0.5
mag, where it should be noted as a slight bump in the luminosity
function of the field stars.
A more generic conclusion is that the presence of a population 
of stars 
similar to those of Arp 2 cannot be excluded in Sgr,  but that, 
in this case, the in-famous HB second
parameter must be invoked.
At this stage we have no indications on the nature of this conjectured
second parameter.

\subsection{Photometry of M54 (NGC6715)}
\par
The study of M54  is important by itself because it lies 
in the densest region of Sgr and, according to Da Costa \& Armandroff (1995),
it is likely to be associated to the Sgr galaxy.
The estimated
metallicity of M54 is [Fe/H]=-1.79$\pm$0.08 (SL),  
very similar to that of Arp 2, but with quite different structural parameters
(central concentrations are 
c=0.90 and c=1.84, respectively for Arp 2 and M54). 

From a simple inspection of the CMD in figure 4a we see that this
 field presents a pronounced  HB, extending bluewards to V-I=0, which,
 as already noted, is completely absent in 
{\it field 1} and {\it field 2} and it is, therefore, to be attributed to M54.
  
The luminosity of the HB was estimated to be V$_{HB}$=18.2$\pm$0.1, 
 having computed the mean and the rms of all the
stars of figure 4a in the box 17.5$\leq$V$\leq$18.5, 0.4$\leq$V-I$\leq$0.8, 
which is likely to
include only HB stars.

Most of the features detectable in fig. 4a 
are naturally present also in the CMD of SL.
In particular, it is visible the  well-populated RGB which extends up to 
V$_{tip}$$\simeq$15.2.  
Some attention deserves the red HB 
located at V$\simeq$18.23 and (V-I)$\simeq$0.85, which is
  fainter than the blue HB. 
In order to identify the nature of this feature we start
trying to estimate the numbers involved,
plotting in figure 11 the luminosity functions of the data of Sgr 
{\it fields 1 and 2} 
and of the {\it field  M54}, for  stars in
the color interval  
0.75$\leq$ V-I $\leq$ 1.0. The counts have been normalized in the 
interval 14.0 $\leq$ V $\leq$ 17.0, and V-I$\leq$ 1.0  
because, given the extremely different
nature of populations of {\it field 1} and {\it field 2} on one side, and
{\it field M54} on the other, we preferred  to refer to the number of 
foreground stars, which are 
largely present in that color and magnitude interval.

%%%%%%% FIG histo comparata dei bumps %%%%%%%%%%%%%%%%%%%%%%%%%%%%%%%%%

\begin{figure}
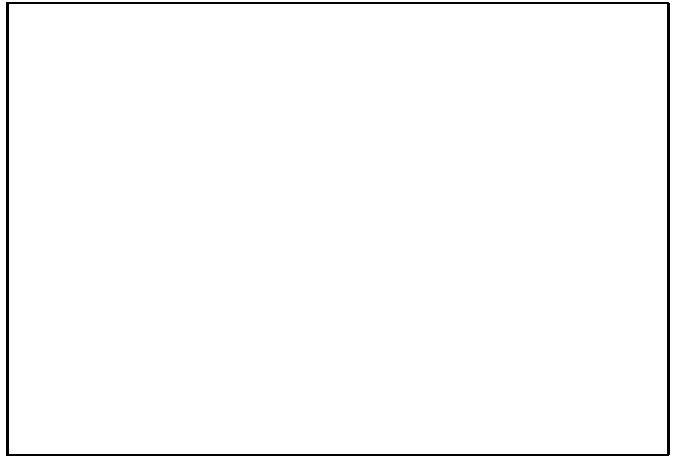


\picplace{6cm}

\caption{Luminosity function of Sgr {\it field 1},
{\it field 2} and of the field centered on M54, in the color range 
0.75$\leq$ V $\leq$ 1.0}

\end{figure}

%%%%%%%%%%%%%%%%%%%%%%%%%%%%%%%%%%%%%%%%%%%%%%%%%%%%%%%%%%%%%%%%%%%%

From figure 11 one immediately 
notes that a bump at V$\simeq$18.09 is clearly 
visible for {\it field 1} and {\it field 2}
while the 
  {\it field M54} presents 
 a more 
prominent bump at V$\simeq$18.23. Therefore, while we 
confirm the early suggestion of SL that such ``anonymous RHB'' belongs 
to Sgr dwarf galaxy, we remain with the problem of explaining 
the different relevance
of the feature in the three fields. 

One (marginally viable) possibility is that we are observing
a simple fluctuation of the data, which show a casual increase 
in that particular color and magnitude interval and in that particular
field.

Another possibility is that such  ``anonymous RHB''  in the {\it field M54}
 is, at least
in part, formed by the photometric blending of stars belonging to the
red HB of the dominant metal-rich population of Sgr and of blue HB
stars of M54. Such a possibility was already suggested in quite a
different context by Ferraro et al. (1991) and its viability is
easy to test. In fact, one can ask whether any ``anonymous RHB star''
could randomly split into two stars, one falling in the region of the
Sgr metal-rich red HB and the other belonging to the blue HB of M54.
Experiments show that this could actually be the case, and 
consequently that the red clump, which apparently forms an extension of
the blue HB of M54 could be, at least in part, a photometric
artifact due to the extremely high crowding of the images. 
       
An objective of obvious interest of the present study 
is to determine the TO luminosity of M54. Given the confusion of 
the TO region, due both to the crowding and 
to the faintness of the images,
we followed 
two independent approaches to the problem. 

First, we divided the turn-off region of figure 4a in intervals of 0.1
magnitudes. Then, we derived for each interval the histogram in color and
fitted these color distribution with a gaussian. The maximum of these gaussian 
 were
finally reported in table 5 in the magnitude interval 20.9$\leq$V$\leq$21.9.
Inspection of table 5 immediately leads to identify the turn-off luminosity of
M54, in correspondence of the 
bluest point of the ridge line
 V$_{TO}$=21.6$\pm$0.1.

%%%%%%%%%%%%%% table %%%%%%%%%%%%%%

\begin {table}
\caption[]{The ridge line of M54 in the turn-off region}
\begin{flushleft}
\begin{tabular}{ccc}
\ \\
\hline
\hline
\ \\
$V$&$(V-I)_{mode}$\\
\hline
\ \\
20.9 & 0.81 & \\
21.0 & 0.77 & \\
21.1 & 0.76 & \\
21.2 & 0.76 & \\
21.3 & 0.76 & \\
21.4 & 0.76 & \\
21.5 & 0.75 & \\
21.6 & 0.73 & \\
21.7 & 0.76 & \\
21.8 & 0.78 & \\
21.9 & 0.79 & \\
\ \\
\hline
\end{tabular}
\end{flushleft}
\end{table}

%%%%%%%%%%%%%%%%%%%%%%%%%%%%%%
   
Having already determined V$_{HB}$ = 18.2$\pm$0.1 one obtains 
$\Delta$V$_{HB}^{TO}$=3.40$\pm$0.14, which, within the errors, 
 is formally identical to
the mean value $\Delta$V$_{HB}^{TO}$=3.55$\pm$0.09 obtained for the
bulk of galactic globular
clusters by Buonanno, Corsi \& Fusi Pecci (1989).

Given the importance of these issue, we approached the problem of determining
 the age of M54 with an alternative 
method based on the morphology of the HB.
The origin of the observed very blue
HB morphologies has been long debated (see Lee, Demarque \& Zinn 1994 
and Fusi Pecci et al. 1996 for two different points of view); however,
introducing a new quantitative observable, (B2-R)/(B+V+R), where B2 is the
number of HB stars with (B-V)$_0$$\leq$-0.02, Buonanno \& Iannicola (1995)
proposed the following relation to compute the relative age of
globular clusters of similar metallicity, taking into account the central
density of the cluster  and the HB
morphology:

\smallskip

$\Delta$t$_9$=3.45(B2-R)/(B+V+R)-0.58log$\rho$$_0$+0.34        (1)

\smallskip

Having adopted for M54 E$_{B-V}$=0.15 and  
log$\rho$$_0$=4.72 (Djorgovski, 1994)
we performed the following counts, B2=19, 
B=100, V=18, R=8, 
obtaining (B2-R)/(B+V+R)=0.09$\pm$0.05, 
 where the
errors are computed on the basis of Poisson statistics
 (figure 12 shows an enlargement of the
HB region of the V$_0$,  (B-V)$_0$ CMD).

The second step was to compare the counts in the M54
to those made in clusters of similar
metallicity and HB morphology. 

%%%%%%%%%%%%%%%%%%%%%%%%%%%%%%%%%%%%%%%%%%%%%%%%%%%%%%%%%%%%%%%%%%%%

\begin{figure}
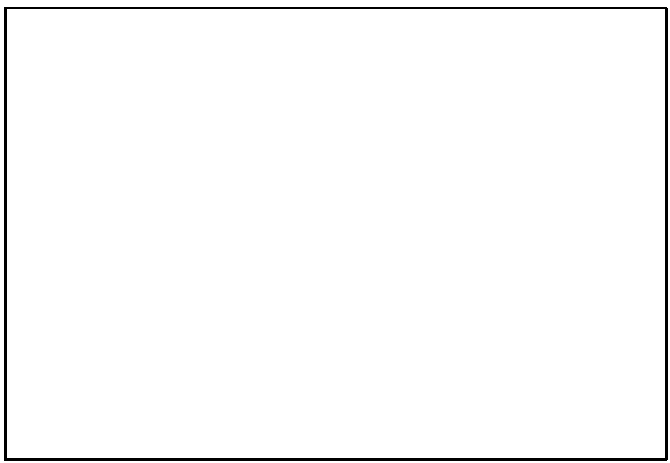


\picplace{6cm}

\caption[]{Enlargement of the HB of M54.
 The four dashed boxes 
are those used for star counts along the HB.}

\end{figure}

%%%%%%%%%%%%%%%%%%%%%%%%%%%%%%%%%%%%%%%%%%%%%%%%%%%%%%%%%%%%%%%%%%%%

We found, in particular, for 
NGC4147 ([Fe/H]=-1.80, log$\rho_0$=3.58)  
(B2-R)/(B+V+R)=0.12$\pm$0.13 (Friel, Heasley \& Christian, 1987), 
and for NGC6101 ([Fe/H]=-1.80$\pm$0.1, 
log$\rho_0$=1.57)   
(B2-R)/(B+V+R)=0.04$\pm$0.02 (Sarajedini \& Da Costa 1991).  

We then applied 
equation (1) differentially, in order to derive the age of M54 relatively to
that of the other two clusters and obtained:

\smallskip

$\Delta$t$_9$(M54-NGC4147) = -0.76 $\pm$0.44 

\smallskip

$\Delta$t$_9$(M54-NGC6101) = -1.63 $\pm$0.17

\smallskip

These relative ages were  then used to pass to {\it relative} 
$\Delta$V$^{TO}_{HB}$ and, finally, to estimate V$_{TO}$ for M54. 
To pass from
relative age to relative TO-HB difference in magnitude, 
we used equation (6b) of  
Buonanno et al. (1989):  

\smallskip

log t$_9$=0.37$\Delta$V$_{HB}^{TO}$-0.06[Fe/H]-0.81, 

\smallskip

and obtained:

\smallskip

$\delta$$\Delta$V$^{TO}_{HB}$(M54-NGC4147)=-0.06$\pm$0.05
% \par\noindent

\smallskip

$\delta$$\Delta$V$^{TO}_{HB}$(M54-NGC6101)=-0.14$\pm$0.04

\smallskip

Adopting for NGC4147 $\Delta$V$^{TO}_{HB}$(4147)=3.60$\pm$0.20 
(Friel et al. 1987) we finally obtained for
M54 $\Delta$V$^{TO}_{HB}$=3.54$\pm$0.21, while adopting for NGC6101 
$\Delta$V$^{TO}_{HB}$(6101)=3.40$\pm$0.20 (Sarajedini \& Da Costa, 1991) 
we obtained for M54
$\Delta$V$^{TO}_{HB}$=3.26$\pm$0.20.  

\smallskip

The average value of the difference in magnitude between the horizontal branch
and the main sequence turn-off for M54 turned out to be:

\smallskip

$\Delta$V$_{HB}^{TO}$=3.40$\pm$0.28

\smallskip\noindent
which 
is in excellent agreement with the
value derived from table 5.

\smallskip\noindent
In conclusion we find that M54, although kinematically 
 associated with Sgr,
 presents the same characteristics (age, HB morphology, second-parameter)
 of a typical galactic globular cluster, and that a M54-like population 
 do not significatively contributes to the Sgr population.

%%%%%%%%%%%%%%%%%%%%%%%%%%%%%%%%%%%%%%%%%%%%%%%%%%%%%%%%%
 
\section{Discussion and summary}
\par

We have performed deep CCD photometry of the Sagittarius dwarf galaxy, 
spanning from 
the main sequence TO 
to the tip of the red giant branch. 

Two of the fields are located in the outskirts of the dwarf galaxy, while the
third field is centered on M54, in such a way that a direct comparison allowed
to single out features which are characteristic of the globular cluster or of
the dwarf galaxy.

The spread in color observed along the RGB of {\it field 1} and {\it field 2}
are beyond any possible photometric errors. The most likely explanation is that
we are observing a spread in metallicity, whose boundaries are those suggested
by the RGB of M2 ([Fe/H]=-1.58) on one side, and the RGB of 47 Tuc
([Fe/H]=-0.71) on the other. In addition, the RGB of the field centered on M54
is even redder than this latter limit, giving hints of the existence of a more
metal-rich population. The presence of an asymptotic giant branch and, then, of
a younger population, cannot be excluded from our data.

The distance modulus has been estimated from the luminosity of the  horizontal 
branch, obtaining 
a value of (m-M)$_0$=16.95, 
corresponding to a distance of d=24.55$\pm$1.0 Kpc, 
in good agreement
with MUSKKK.

Probably the most important result of the present study is that comparing the
CMD of Sgr with those of
the kinematically associated globulars, we found that the dominant stellar 
population of Sgr is very similar to the population 
of the ``young'' globular 
Ter 7, while
the presence of a population 
 Arp2-like is dubious. The population of M54 appears clearly  different
from that of Sgr. All these findings
put strong constraints to the Sgr star formation history.    

%%%%%%%%%%%%%%% fine %%%%%%%%%%%%%%%%%%

\begin{acknowledgements} 
We thank M. Limongi, A. Tornambe', 
and S. Cassisi for 
helpfull comments and suggestions.

\end{acknowledgements}

\end{document}